# Astronomical Anomalies: Their Role in the Quest for Extraterrestrial Life

Beatriz Villarroel & Geoffrey W. Marcy

## Abstract

Astronomers occasionally detect an object having unexpected shape, unexplainable photometry, or unprecedented spectra that are inconsistent with our contemporary knowledge of the universe. Upon careful assessment, many of these anomalies are discarded as mere noise, contamination, or faulty analysis. But some anomalies survive scrutiny to yield new astronomical objects and physical processes. Examples of validated anomalies include quasars, pulsars, and periodic Doppler shifts of Sun-like stars caused by the gravitational pull of orbiting planets. Other anomalies persist as mysteries, including Fast Radio Bursts, dark energy, 'Oumuamua as an alien spaceship, and simultaneously vanishing stars. Advanced technological life may present astronomers with anomalies that require carefully designed observations from multiple vantage points simultaneously and with real-time spectroscopy.

## Great Anomalous Moments in Astronomy

The word "anomaly" is defined by the Oxford English Dictionary as "a thing, situation, etc. that is different from what is normal or expected." In the field of astronomy, an anomaly becomes the subject of detailed observations, studied with images, spectra, and light curves. Sometimes, once the anomaly has been discovered and published in a reputable peer reviewed journal, fellow colleagues pick up on it and decide to perform their own analysis.

When Maarten Schmidt spotted a bright dot of light with a greatly Doppler shifted spectrum in 1963, the object was considered an anomaly. Within a few years, further observations showed these objects, now known as quasars, are actually outside our own Galaxy. Schmidt is credited with the discovery of the first quasar, known as 3c 273 (Schmidt, 1963). Twenty years later, astronomers realized that quasars are luminous centers of galaxies powered by hot gas falling onto supermassive black holes. Astronomers have now cataloged millions of them. Observations show that super massive black holes exist at the centers of nearly all galaxies, all of which were "active galactic nuclei" in the past when accreting gas. Thus, we now have a complete turn-around: A galaxy that does not harbor a supermassive black hole and was never a quasar is now the anomaly.

History shows that an anomaly is at risk of being discarded as the result of a faulty analysis. Other times, the anomaly makes a comeback after having been buried, forgotten, or dismissed due to biases, sociological factors, or the passage of time. One still unresolved anomaly is Halton Arp's famous discovery of *anomalous redshifts*, where pairs of galaxies apparently connected by bridges nonetheless have different redshifts. According to the commonly accepted expanding universe model, the redshifts indicate vastly different distances for the two galaxies, which would conflict with the apparent "bridge" between the two. Arp and others, including anti-Big Bang-proponents Margaret and Geoffrey Burbidge and Fred Hoyle, soon found many more examples of galaxies with anomalous redshifts, claiming that quasars tended to surround themselves with many more background and foreground objects than normal galaxies (Hoyle & Burbidge, 1996, Burbidge et al., 2003). These anomalous bridged but redshift-discrepant quasars challenged the notion that the redshift of a quasar truly indicates its distance from us. Thus, the Universe might not be expanding.

When the Armenian astronomer Victor Ambartsumian hypothesized that galaxies are actually splitting and giving birth to one another, rather than gravitationally colliding, Halton Arp proposed that quasars actually are newly born galaxies ejected from a galaxy core (obtaining a high redshift upon ejection), and thus could be found in alignment with other objects. After a few heated debates, the discussion of anomalous redshifts died off without a final resolution (López-Corredoira, 2010) even as the consensus has been to disregard the effects as due to poor statistical analysis. So an anomaly might be abandoned before it's even solved, simply because scientists find no meaningful resolution and grow tired of it! Each subtopic of modern astronomical research has its own set of anomalies.

Today, some of the most fascinating anomalies are those that relate to our searches for extraterrestrial life. Some of these anomalies have been thoroughly reported in the popular science media. The question of whether or not we humans are alone as a technological civilization within our Milky Way Galaxy is so intriguing to scientists and the public that even a remote possibility that an anomalous observation is caused by the activity or presence of "little green men" serves as a driving force to find an underlying explanation.

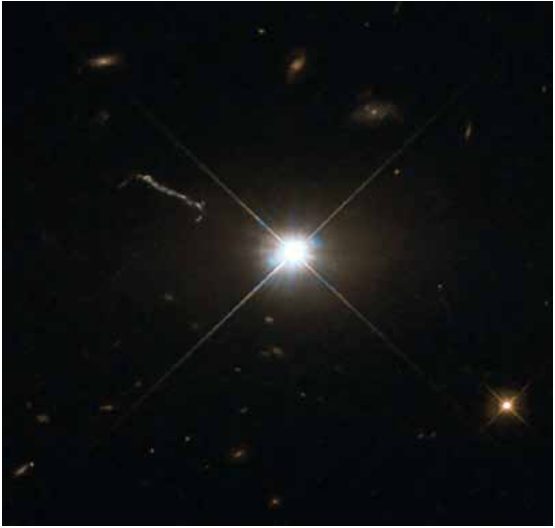

*Figure 1. Quasar 3c 273 seen at optical wavelengths. (ESA/Hubble & NASA)*

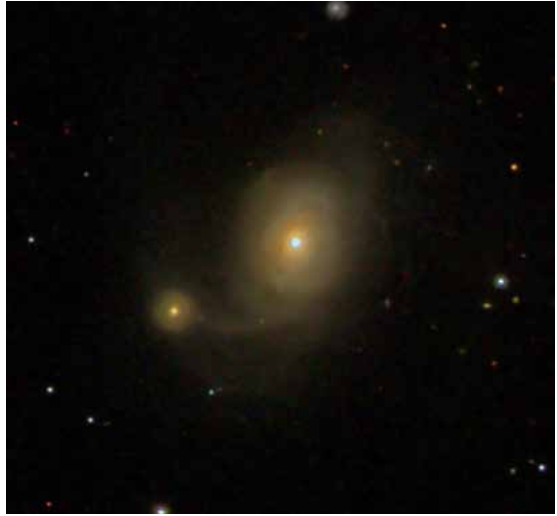

*Figure 2. NGC 7603 contains a pair of galaxies connected by an apparent "bridge". The galaxies are located a two very different redshifts, indicating very different distances. (Sloan Digital Sky Survey)*

## From Little Green Men to Exoplanets

In 1959, Giuseppe Cocconi and Philip Morrison published a seminal work, "Searching for Interstellar Communications," in the journal Nature where they speculated that just as we humans are unwittingly broadcasting our existence to our interstellar neighbors through radio waves, our interstellar neighbors may similarly be revealing their presence to us. When a young PhD student named Jocelyn Bell, one late summer in 1967 in rainy Cambridge, detected a mysterious repeating radio signal (Hewish et al., 1968) with a new telescope she herself had helped build, it was no surprise that she immediately alerted her advisor, Anthony Hewish. They found the radio waves arrived in pulses, each lasting a fraction of a second, followed by a pause of 1.3 seconds. The mysterious radio source came from a particular coordinate in the sky, which made it obvious that it did not come from our own Earth. What could this mysterious source be?

Soon after this signal was referred to as "Little Green Men-1" (LGM-1), and astronomers started considering the possibility of alien activity. This anomaly would have remained mysterious if others had not been found, making it clear that pulsating objects are a natural phenomenon and abundant in the universe. Today, pulsars are understood to be dense, compact objects only a few tens of kilometers in size, with a strong magnetic field (up to a trillion Gauss), and that spin hundreds of times each second around their axis, emitting double beams of accelerated particles that we observe on and off as the object quickly rotates. These quizzical false-alarm little green men form at the end of the lives of massive stars, when they explode as supernovae only to collapse to a neutron star. Anthony Hewish went on to receive the Nobel Prize for the discovery in 1974, while Jocelyn Bell Burnell had to wait 44 more years to receive the nearly as prestigious Special Breakthrough Prize in Fundamental Physics in 2018.

But that was not the end of the story with pulsars. Their regular pulses constitute accurate clocks and they sometimes come in orbiting pairs bound by gravity. So these "orbiting clocks" provide an excellent test of

Einstein's theory that predicts the existence of gravitational waves. When one such pair was discovered, gravitational waves were demonstrated indirectly through the decrease of the orbital period of the system as the two neutron stars spiral toward each other. The pair of pulsars was losing energy through the emission of gravitational waves, as Einstein's theory predicted. Impressed by the agreement between the predictions and the observations (Weisberg et al., 1981, Taylor & Weisberg, 1982), the Swedish Royal Academy of Sciences decided to award a second Nobel Prize for pulsars, this time going to Russell Hulse and Joseph Taylor Jr.

In 1992, Alexander Wolszczan and Dave Frail announced the discovery of the first exoplanet found outside the Solar System (Wolszczan & Frail, 1992). This was the confirmation that other star systems might form planets as well, not only those that formed around our own Sun. A few years later, the first three exoplanets around normal, hydrogen-burning stars were found by two competing teams, Michel Mayor and Didier Queloz (1995) and Geoffrey Marcy and Paul Butler (1996), accomplished by detecting Doppler shifts that vary with a periodicity.

However, the first periodic Doppler shift found around 51 Pegasi was widely doubted as due to an actual planet, and for good reasons. The orbital period was anomalously short, only 4.3 days, much shorter than any "normal" planet in our Solar System. Also, the Doppler shift varied as a perfect sine wave, which is consistent with a star that was merely pulsating, breathing in and out. Such pulsations are well known to exist and constituted a less "anomalous" explanation than the existence of a "planet." Moreover, the periodic Doppler shifts could result from a binary star in a nearly "face-on" orbit. However, the discovery of Doppler periodicity in the Sun-like star 70 Virginis was definitive due to a planet. Its period of 117 days placed it at a distance comparable to Mercury's orbit. Even better, the orbit was clearly eccentric, with an oval shape exactly as Newton's Laws predicted, making it unmistakably an orbiting planet rather than the pulsation of star (Marcy & Butler, 1996). 70 Virginis, along with 47 Ursae Majoris with its two-year orbital period, pointed to a high prevalence of planetary systems in a diversity of orbits, and many opportunities for life.

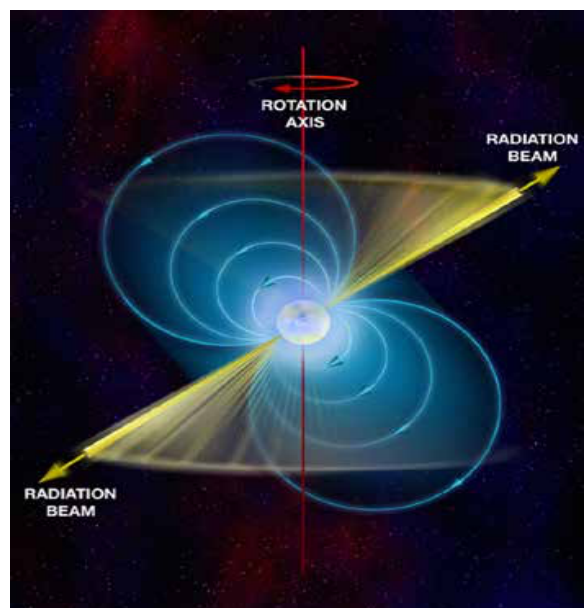

*Figure 3. A sketch of the structure of a rotating pulsar that was initially anomalous but now serves as an accurate cosmic clock. (courtesy, Bill Saxton/NRAO/AUI/NSF)*

The Discovery of Fast Radio Bursts

About 15 years ago, two researchers at the Parkes Observatory were scanning archival data taken some years earlier with the radio telescope. In the data, they saw a short burst, lasting less than 5 milliseconds. The short burst was extremely powerful and came from a source outside our galaxy. A new anomaly in the radio wavelengths had been found. This first Fast Radio Burst (FRB) was discovered by Duncan Lorimer and his student David Narkevic (Lorimer et al., 2007). Today, we know of about 500 FRBs, some repeating themselves irregularly, a few regularly. The spacing between the pulses varies, and some can have periods as long as 18 minutes (see e.g. Hurley-Walker et al., 2022). The FRBs are emitted from a source with an extremely strong magnetic field. The origin of these bursts is heavily debated; they are thought to be caused either by very magnetized neutron stars or possibly by the merger of compact objects. It is difficult to build reliable theoretical models of these objects because these bursts are very short and difficult to localize in the radio spectrum. Last year some significant progress was made when a FRB inside our own Galaxy, FRB 200428, was localized by the Canadian Hydrogen Intensity Mapping Experiment (The CHIME/FRB Collaboration, 2020). The FRB was also emitting in the gamma and x-ray frequencies. The scientists proposed that the burst of light could be produced when the jet from a magnetar (a neutron star with an extremely strong magnetic field) collides with the surrounding interstellar medium and produces a shock wave.

But even here astronomers have not refrained from discussing alternative possibilities such as ET technology (Lingam & Loeb, 2017). The speculations on the possible ET origin of FRBs peaked at a time when a particular subcategory of FRBs was found and dubbed "perytons," radio bursts having pulses that were about 250 milliseconds in duration at the 1.4 GHz radio frequency (Burke-Spolaor et al., 2011). Was ET signaling to us via these very peculiar and sharp bursts? The mystery further deepened as researchers noticed that the bursts only appeared weekdays and during office hours. How clever—ET must already be familiar with our habits! More clues emerged until it was finally established that the perytons were caused by two microwave ovens at the observatory that emitted small bursts whenever the lunch goer opened the microwave door prematurely (Petroff et al., 2015)!

## A Megastructure Around Boyajian's Star

Sometimes an anomaly is an error or an instrumental fluke, but in some cases the object actually has an anomalous behavior that might or might not be natural in its origin. An example is Boyajian's Star named after its discoverer Tabetha Boyajian (Boyajian et al. 2016). The object showed an unusual dimming. The discovery hit the headlines of all major media in 2016. Jason Wright et al (2016) described their extraordinary theory for the dimmings of the star this way: "KIC 846285 [is] an object with a bizarre light curve consistent with a 'swarm' of megastructures. We suggest that this is an outstanding SETI target."

The news media exaggerated this claim in hundreds of newspapers, blogs, podcasts, and documentaries, all highlighting megastructures built by alien super civilizations, with little mention of the heritage explanation: clumps of dust that could block a small fraction of Boyajian's star. The media seldom report the more conventional interpretations. It was well known since the 1940s that young Sun-like stars, called T Tauri stars, vary in brightness sporadically and unpredictably due to dust clouds around them that occasionally move in front of the star, blocking some of the star's light (Bertout, 1989). Indeed, stars somewhat older than our Sun also show brightness variations due to dust that hasn't gone away over the millions of years. Even our Sun still has some dust around it; it's called "Zodiacal dust" and it's the remains of collisions of asteroids that fragment into dust. But in the end, thanks to the large interest in the story, extensive measurements were made of its brightness at different wavelengths from blue to red to the infrared, which showed colors consistent with dust along the line of sight—and the anomaly of Boyajian's Star has gone away, for now.

## There is Life in Venus Atmosphere

Even when the news reporting is more balanced, not every anomaly survives the test of time. One such example was the suggested discovery of life on Venus last year (Greaves et al., 2021). The discovery was intriguing because we are unaware of any organisms that can survive in a climate as hot as Venus, which has a surface temperature of 475 C. The scientists had analyzed radio waves coming from Venus using the James Clerk Maxwell Telescope and Atacama Large Millimeter/submillimeter Array. They reported absorption of wavelengths at which phosphine absorbs radio waves, thus discovering the gas phosphine in the atmosphere of Venus. Theorists in a group headed by Sara Seager developed atmospheric and chemical models related to microbial life in a "Venusian Aerial Biosphere" (Seager et al., 2021), and their theoretical models predicted that phosphine could not possibly occur unless some microbial life form generated it. So they concluded that the discovery of phosphine constituted strong evidence for life in the atmosphere of Venus.

Doubts were soon raised, however. One was related to the measurement technique and the method used to identify and quantify the phosphine in the spectrum. Another was that even if the signal was real, the absorption could be caused by some other molecule. Indeed, sulfur dioxide absorbs at nearly the same wavelength (Lincowski et al., 2021). Life in the atmosphere of Venus also seems unlikely because the clouds are composed of sulfuric acid, and also any descending organism would fall onto the cauldron of the surface. Even if the signal truly came from phosphine, non-biological origins remain possible, including volcanism (Truong & Lunine, 2021, Bains et al., 2022). Three follow-up observations, including at radio and infrared wavelengths, failed to find phosphine (Snellen et al., 2020; Villanueva et al., 2021; Lincowski et al., 2021). In this case, we see that an interested community can help to quickly investigate the most intriguing claims. But even these latest results are subject to change as some recent indications suggest the presence of ammonia in droplets in Venus atmosphere, making the planet more habitable than previously thought. Not seldom, it takes years to reach a final conclusion and it may well be the case here as well. We must remain open to the possibility that the phosphine or life might still be detected on Venus.

## An Alien Spaceship Inside the Solar System

The most heated anomaly since Halton Arp's discrepant redshifts may be 'Oumuamua, which was discovered in 2017 as a point of light moving through the night sky. It entered our Solar System from outside, and then exited, never to return. It was the first object ever discovered to enter our Solar System from outside. It passed through unexpectedly and too quickly to allow careful observations, leaving its properties poorly measured (Meech et al., 2017; ´Cuk, 2018; Raymond et al., 2018; Jewitt & Luu, 2019; Moro-Martin 2019). Although first thought to be comet, it had no cometary tail, and unlike other comets showed no carbon-based molecules or dust (Trilling et al., 2018). Extensive observations showed that its speed was too great to remain bound to the Sun (JPL, 2017).

A debate about its nature took off. Asteroids and comets are commonly "sling-shot" by gravity as they pass near planets, often achieving escape velocity from their home planetary system. As planetary systems are common around stars, our Milky Way Galaxy must contain millions of billions ($> 10^{15}$) of these wandering rocky escapees, some of which will pass through our Solar System by chance. Thus, 'Oumuamua has a natural explanation, albeit with some observational properties that remain unresolved (e.g. Jewitt et al., 2017; Meech et al., 2017; Luu et al., 2019; Moro-Martin, 2019).

But the natural explanation was not shared by all scientists. Shmuel Bialy and Abraham Loeb (2018) proposed that 'Oumuamua is a spacecraft with a light sail constructed by an advanced civilization. The light sail would be thin, less than a millimeter in thickness, and large enough to allow the object to be pushed by the reflection of sunlight. The extra push on the motion of the comet could be explained with a light sail. This suggestion gained international attention. But many scientists dismiss this spacecraft explanation, given the existence of possible natural explanations.

There is certainly a common "bias" against the spacecraft explanation for 'Oumuamua among scientists. The bias has several origins. First, we humans have never detected technological artifacts in the Galaxy, constituting an "absence of evidence" of alien life despite a century of telescopic observations. Second, the spacecraft explanation attracts too much media attention, which makes some scientists uncomfortable. Third, dogmatic ideas are as prevalent in the scientific community as in any other human endeavor (López-Corredoira & Castro Perelman, 2008).

But the spacecraft theory of 'Oumuamua deserves an assessment of our bias against it. Suppose we had prior information about the prevalence of technological civilizations in our Galaxy, such as observations of radio or laser beacons detected from many directions. Such information might change our "prior" assumptions about the possibility of spacecraft and light sails, functional or not. The relative probability that 'Oumuamua is of natural versus technological origin depends on our prior information, motivating caution about our "gut feeling."

In September 2020, another interstellar visitor with similar properties as 'Oumuamua was detected, named 2020 SO. This visitor lacked any signs of outgassing. It turned out to be a rocket booster from a 1966 mission to the Moon.

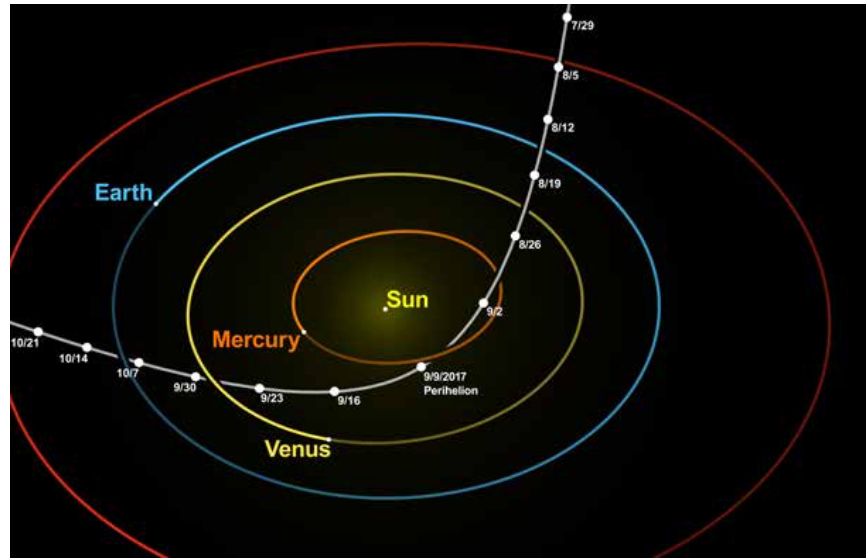

*Figure 4. Trajectory of 'Oumuamua through the inner parts of the Solar System in 2017, dated weekly. (Credit: Avi Loeb and Nagual Design, Creative Commons Attribution 4.0 International)*

## Searching for "Vanishing Stars" with the VASCO Project

So far we have seen examples of serendipitous discoveries of anomalies. But one can also search for anomalies deliberately. An example of a project that aims to find weird astrophysical transients in the sky is the Vanishing & Appearing Sources during a Century of Observations (VASCO) project. The project proposal made the case for looking for "vanishing stars" and "impossible effects'' in an attempt to find (1) signs of massive, evolved stars that fail to emit a bright supernova as they collapse to a black hole, (2) the activity of extraterrestrial intelligence in action through, for example, interstellar beacons, and (3) new, speculative phenomena like wormholes (Villarroel et al., 2016). From its inception, the project took on the goal of comparing historical images of the sky in the 1950s with modern sky imagery from Pan-STARRS observatory. An example of a transient that is visible in the old Palomar images is the object 1084-0241525 in the USNO-B1.0 catalog. Deep new observations with the 2.5m Nordic Optical Telescope in the Canary Islands eventually revealed a counterpart to the star, suggesting that the object was something that flared up momentarily in the old images. Soon, the VASCO project found about a hundred short-lived transients that were visible point sources in the old Palomar images but not to be seen in PanSTARRS data. Most of these are thought to be natural transients, such as flaring stars, optical counterparts to gamma-ray bursts, and similar phenomena. A list of such short-lived transients can even be useful in searches for communication lasers (Villarroel et al., 2020) that also would leave bright, short-lived spots in an image.

Last year, an unusual discovery of nine simultaneously appearing and vanishing star-like objects in a small image—a small fraction of a square degree in the sky—was announced (Villarroel, Marcy, Geier et al., 2021). An image taken half an hour earlier, and another image taken six days later, had no transients in the spots, suggesting that they appeared and vanished within the exposure time of the plate. No astrophysical scenario could be reconciled with this finding. A number of tests for instrumental artifacts also failed, revealing nothing dubious about the nine spots. Finally, the authors proposed that the "nine simultaneous transients" could be some "unknown" type of photographic plate contamination (defects) that produced eerily star-like shapes of varying intensities. But they also mentioned the possibility of having found a new phenomenon on the sky. One possibility discussed was reflections from highly reflective and flat objects in orbits around the Earth. To test this idea further, the VASCO team recently proposed how to look for solar reflections from objects in pre-satellite images (Villarroel et al., 2022) by looking for images where multiple transients follow a line. In this case, whether one finds support for the hypothesis or not, the result will be valuable for studies of possible alien artifacts in orbit around the Earth and to estimate an upper limit of such hypothetical objects. But a solution to the simultaneous transients might be neither plate defects nor alien artifacts. Nature tends to surprise us continuously with new phenomena, and maybe also here, the explanation might be rooted in physics presently unknown to us.

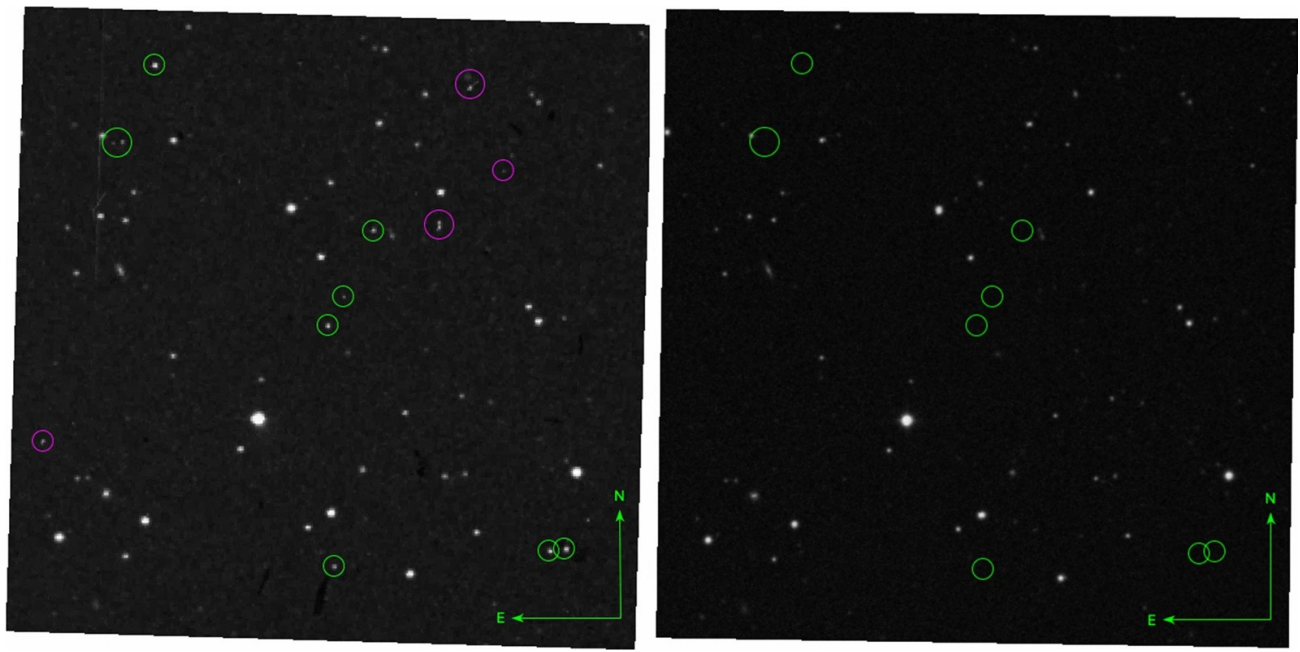

*Figure 5. Nine simultaneous transients. The green circles show nine star-like objects seen in a red photographic plate image of the sky taken on April 12, 1950 (left) but not seen in an image of the same patch of sky taken 1996 (right). Comparison with other photographic plates shows that these objects appear and vanish within the exposure time of the plate, thus "simultaneously". (Purple circles are artifacts during the scanning process.)*

## Progress and Possibility

The discovery of anomalies offers progress in astronomy through discrepancies with expectations, curiosity, and debate. The "little green men" gave researchers an extra push needed to spark human imagination and a desire to gather more observations. Many scientists quote the loss of "credibility" of a certain field when big claims of alien life are made to the media. On the other hand, we have shown with several examples how research activity is often stimulated by the possibility of discovering alien life, regardless of the outcome.

The authors wish to thank Martin J. Ward for helpful discussions.